\documentclass{aa}
\usepackage{graphicx}
\usepackage{txfonts}
\usepackage{multirow}
\usepackage{url}

\begin{document} 

   \title{Deep learning for Sunyaev-Zel'dovich detection in \textit{Planck}}

   \author{V. Bonjean\inst{1, 2}
          }

   \institute{Institut d'Astrophysique Spatiale, CNRS, Université Paris-Sud, Bâtiment 121, Orsay, France\\
   \email{victor.bonjean@ias.u-psud.fr}
         \and
             LERMA, Observatoire de Paris, PSL Research University, CNRS, Sorbonne Universités, UPMC Univ. Paris 06, 75014, Paris, France\\
             }

   \date{Received XXX; accepted XXX}

  \abstract
{
The \textit{Planck} collaboration has extensively used the six \textit{Planck} HFI frequency maps to detect the Sunyaev-Zel'dovich (SZ) effect with dedicated methods, e.g., by applying (i) component separation to construct a full sky map of the $y$ parameter or (ii) matched multi-filters to detect galaxy clusters via their hot gas. Although powerful, these methods may still introduce biases in the detection of the sources or in the reconstruction of the SZ signal due to prior knowledge (e.g., the use of the GNFW profile model as a proxy for the shape of galaxy clusters, which is accurate on average but not on individual clusters). In this study, we use deep learning algorithms, more specifically a U-Net architecture network, to detect the SZ signal from the \textit{Planck} HFI frequency maps. The U-Net shows very good performance, recovering the \textit{Planck} clusters in a test area. In the full sky, \textit{Planck} clusters are also recovered, together with more than 18,000 other potential SZ sources, for which we have statistical hints of galaxy cluster signatures by stacking at their positions several full sky maps at different wavelengths (i.e., the CMB lensing map from \textit{Planck}, maps of galaxy over-densities, and the ROSAT X-ray map). The diffuse SZ emission is also recovered around known large-scale structures such as Shapley, A399-A401, Coma, and Leo. Results shown in this proof-of-concept study are promising for potential future detection of galaxy clusters with low SZ pressure with this kind of approach, and more generally for potential identification and characterisation of large-scale structures of the Universe via their hot gas.
}


   \maketitle

\section{Introduction}

In the last decades, new statistical developments have begun to play an important role in data reduction and in data analysis. Particularly, the studies involving machine learning algorithms have increased exponentially, as they are very efficient to identify commonalities in a large amount of data, as well as detecting very faint and/or very complex patterns. In the machine learning domain, there are two main families of algorithms: the unsupervised, and the supervised ones. In the first case, algorithms are designed to work on unlabelled data. This is the case for some clustering algorithms such as k-means \citep{macqueen1967}, soft k-means \citep{bezdek1981}, or Gaussian Mixture Models \citep[GMM,][]{dempster1977}, for density estimation algorithms such as Generative Adversial Networks \citep[GAN,][]{goodfellow2014}, or for dimensionality reduction algorithms such as Autoencoders \citep{kramer1991}, Self-Organizing Maps \citep[SOM,][]{kohonen1982}, or principal curves and manifold learning \citep{hastie1989}. In the second case, machine learning algorithms are designed to estimate properties or labels, based on inputs and outputs, both provided by the user. The user must in this case have a perfect knowledge of the labels or of the properties of reference used as output in the training catalogue. This type includes algorithms such as Artificial Neural Networks \citep[ANN,][]{neuralnetwork}, Random Forests \citep[RF,][]{randomforest}, Support Vector Machine \citep[SVM,][]{svm}. Some algorithms, with very complex architectures of superposed layers, may enter in the category of Deep Learning (DL) algorithms; as this is the case for ANN and for Convolutional Neural Networks \citep[CNN,][]{cnn}.

Machine learning algorithms have already been applied successfully in astronomy, astrophysics, and cosmology \citep[e.g.,][for a review]{baron2019}. For example, unsupervised algorithms are used to reconstruct the Cosmic Web (e.g., Bonnaire et al., in prep.), while supervised machine learning algorithms, like ANN or RF, have been used to estimate galaxy redshifts or galaxy types \citep[e.g.,][]{bilicki2014, bilicki2016, krakowski2016, siudek2018, bonjean2019}, to estimate spectral properties of sources \citep[e.g.,][]{ucci2018}, to classify sources \citep[e.g.,][]{aghanim2015}, to search for variable stars \citep[e.g.,][]{pashchenko2018}, as a very non-exhaustive list of examples of applications. More sophisticated algorithms of machine learning, like DL algorithms, widely improve the results compared to those obtained with physical models. In most cases, the computation time required to estimate the results is also significantly reduced. For instance, DL algorithms have been already used to estimate galaxy morphologies and redshifts \citep[e.g.,][]{huertas2015, pasquet2019, boucaud2019}, to fit galaxy surface brightness profiles \citep[e.g.,][]{tuccillo2018}, to compare galaxy surveys \citep[e.g.,][]{dominguez2019}, to detect cosmic structures \citep[e.g.,][]{aragon2019}, to learn the structure formation from initial conditions \citep[e.g.,][]{lucie2018, he2018}, or to generate fast Cosmic Web simulations \citep[e.g.,][Ullmo et al., in prep.]{rodriguez2018}. For galaxy cluster studies only, ML algorithms have been extensively used to successfully compute their properties such as their masses \citep[e.g.,][]{ntampaka2015, ntampaka2016, green2019, calderon2019, ho2019}. Therefore, those very powerful algorithms may help us in the near future to deal with the huge amount of data the community is collecting, and to answer some open questions in astrophysics and cosmology.

Today, the quest of the missing baryons remains one of the biggest challenge in matter of cosmology \citep{fukugita1998, cen1999, shull2012, degraaff2019}. They are expected to hide in the cosmic filaments between the nodes of the cosmic web (i.e., the galaxy clusters), in a form of a warm hot inter-galactic medium (WHIM), heated to a temperature around $10^5<T<10^7$ K. These baryons are in a phase just too hot to be detectable by molecular gas tracers, and not hot enough to be detectable in X-rays.

The Sunyaev-Zel'dovich effect \citep[SZ,][]{sunyaev1970, sunyaev1972}, that is the inverse Compton scattering by the free electrons in the hot ionised gas that redistributes the energies of the cosmic microwave background (CMB) photons, is an ideal candidate to chase the WHIM. This effect is quantifiable by the Compton parameter, $y$:

\begin{equation}
y = \frac{\sigma_\mathrm{T}}{m_\mathrm{e}c^2}\int n_\mathrm{e}(l)k_\mathrm{B}T_\mathrm{e}(l)\mathrm{d}l,
\end{equation}

where $\sigma_\mathrm{T}$ is the Thomson cross-section, $m_{\mathrm{e}}$ the mass of the electron, $c$ the speed of light, $k_{\mathrm{B}}$ the Boltzmann constant, and $n_{\mathrm{e}}(l)$ and $T_{\mathrm{e}}(l)$ being the density and the temperature of the free electrons along the line of sight, respectively. Proportional to $n_\mathrm{e}\times T$, the SZ emission is more sensitive to lower pressure gas, compare to the X-ray emission from Bremsstrahlung process that is sensitive to ${n_\mathrm{e}}^2$.

\begin{figure*}[!ht]
\centering
\includegraphics[width=0.5\textwidth]{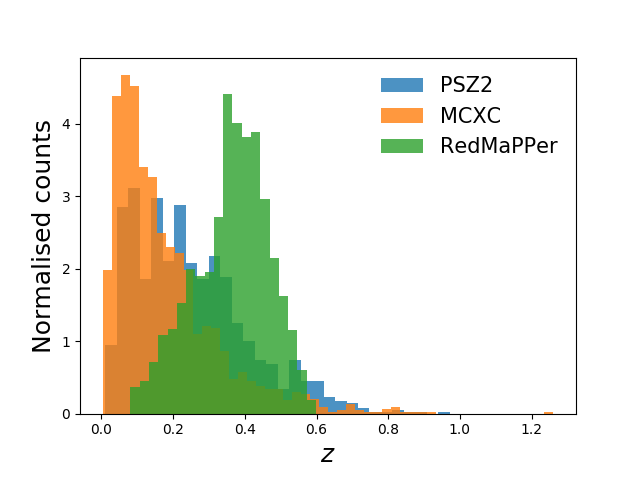}\includegraphics[width=0.5\textwidth]{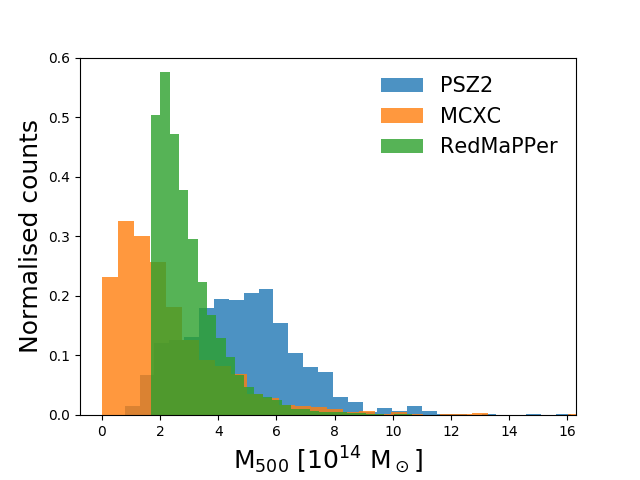}
\caption{\label{cluster_prop}Redshift $z$ and mass $\mathrm{M}_{500}$ distributions of the three galaxy cluster catalogues used in this study: the PSZ2, the MCXC, and the RedMaPPer catalogue.}
\end{figure*}

The \textit{Planck} satellite \citep{tauber2010}, thanks to its spectral coverage between 30 GHz and 857 GHz, has provided ideal data to capture the signature of the SZ effect, which results in a negative emission as compare to the CMB spectrum below 217 GHz, and a positive emission beyond. Based on two component separation techniques, i.e., the Needlet Internal Linear Combination \citep[NILC,][]{remazeilles2011} and the Modified Internal Linear Combination Algorithm \citep[MILCA,][]{hurier2013}, the \textit{Planck} collaboration has constructed full-sky maps of the $y$ SZ Compton parameter at a resolution of 10 arcmin, using the six frequencies of HFI \citep{planck_psz22016}. As a matter of fact, the first statistical detections of the WHIM in its hottest phase, and of part of the missing baryons at low redshift were made using the SZ maps from \textit{Planck} \citep{tanimura2019, tanimura_sc2019, degraaff2019}. The \textit{Planck} collaboration has also applied Matched Multi-Filters on the six \textit{Planck} HFI frequency maps to detect hundreds of new galaxy clusters via the SZ effect, that were later confirmed in optical \citep[e.g.,][]{planck_psz22016, streblyanska2019}. Since then, some studies have shown promising results by increasing the number of \textit{Planck} SZ cluster sources to about 3500 by using new approaches, like artificial neural networks \citep{hurier2017}, or by combining with other wavelengths, e.g., in X-ray with ROSAT \citep{tarrio2019}. Other studies have aimed at producing new higher resolution and lower noise SZ maps by combining \textit{Planck} and ACT data \citep{aghanim2019}. New detections of individual clusters or of stacked diffuse gas are still ongoing, showing that the full potential of the \textit{Planck} data has not been completely exploited. 

Studies deriving SZ catalogues or maps have combined the data with prior knowledge (e.g., assuming a Generalised Navarro, Frenk, and White \citep[GNFW,][]{nagai2007, arnaud2010} profile), or by degrading the resolutions to the highest angular beam for homogeneity \citep{planck_sz2016}. Those biased knowledge priors may prevent us from using the full potential of the \textit{Planck} data. In this study, we propose the application of deep learning algorithms on the \textit{Planck} data to detect low signal-to-noise SZ sources by training on high signal-to-noise SZ sources, i.e., galaxy clusters. We present the data used for the training in Sect.~\ref{sect:data}, and the learning procedure in Sect.~\ref{sect:methods}. In Sect.~\ref{sect:results} and Sect.~\ref{sect:whim}, we present the different results, and we summary the work in Sect.~\ref{sect:discussion}.

\section{Data}\label{sect:data}

\subsection{\textit{Planck} maps}\label{sect:data_map}

\textit{Planck} is the third generation of satellite that aimed at studying the CMB, after COBE \citep{boggess1992} and WMAP \citep{mather1990}. \textit{Planck} comprised two main instruments: the Low Frequency Instrument \citep[LFI,][]{bersanelli2010, mennella2011}, which has observed the sky at 30, 44, and 70 GHz with angular resolutions respectively of 33.29, 27.00, and 13.21 arcmin, and the High Frequency Instrument \citep[HFI,][]{lamarre2010, planck_hfi2011} that observed the sky at 100, 143, 217, 353, 545, and 857 GHz with angular resolutions of 9.68, 7.30,5.02, 4.94, 4.83, and 4.64 arcmin, respectively.

The \textit{Planck} collaboration has provided the community with nine full-sky maps of the sky, all of them publicly available on the Planck Legacy Archive\footnote{\url{https://pla.esac.esa.int/#home}}, in the HEALPIX format \citep{gorski2005}. Considering the increase of the beam with the decrease of the frequency (especially in the frequencies of LFI), and considering that HFI frequencies encompass well the spectrum of the SZ effect, we have decided to work only with the six maps in the frequencies of HFI, i.e., from 100 to 857 GHz.

We also used in this study the latest MILCA \textit{Planck} $y$ map from 2015 \cite{planck_sz2016}. This map is publicly available\footnote{\url{http://pla.esac.esa.int/pla/}} in the HEALPIX format with $n_{\mathrm{side}}$=2048 and a pixel size of $\theta_{\mathrm{pix}}$=1.7 arcmin.

\subsection{PSZ2 cluster catalogue}

The \textit{Planck} collaboration has used the six \textit{Planck} HFI frequency maps with multi-match filters, filtering with GNFW pressure profile model, and taking into account the beam at each frequencies and the spectral dependency of the SZ effect. They have implemented three different algorithms for the cluster detection: two implementations of the Matched Multi-Filter (MMF1 \citep{herranz2002} and MMF3 \citep{melin2006}), and PowellSnakes \citep[PwS][]{carvalho2009, carvalho2012}. They have detected, with the union of the three methods, 1,653 galaxy cluster candidates with a signal-to-noise ratio greater than 4.5 $\sigma$ \citep{planck_psz22016}. The purity of the catalogue is of 83\% \citep{planck_psz22016}, leading to about 300 false detections, which are infrared or CO residual sources. For the confirmed galaxy clusters with measured redshifts, mass $\mathrm{M}_{500}$ are provided in the catalogue, estimated following \cite{planck_PSZ12014}. The mass and redshift distributions of the PSZ2 clusters with confirmed redshifts are shown in Fig.~\ref{cluster_prop}.

\subsection{MCXC cluster catalogue}

Galaxy clusters can also be detected via the hot gas in the X-rays through Bremsstrahlung emission. The ROSAT All-Sky Survey \citep[RASS,][]{truemper1982} is to date the only full-sky survey in X-rays. Galaxy clusters detected based on ROSAT were combined to build a meta-catalogue: the Meta-Catalogue of X-ray detected Clusters \citep[MCXC,][]{piffaretti2011}. The MCXC combines galaxy clusters from RASS-based catalogues (i.e., the Northern ROSAT All-Sky Survey \citep[NORAS,][]{boehringer2000}, the ROSAT-ESO Flux Limited X-ray Survey \citep[REFLEX,][]{boehringer2004}, the ROSAT brightest cluster sample \citep[BCS,][]{ebeling1998}, galaxy clusters around the South Galactic Pole \citep[SGP,][]{cruddace2002}, galaxy clusters around the North Ecliptic Pole \citep[NEP,][]{henry2006}, the Massive Cluster Survey \citep[MACS,][]{ebeling2001}, and the Clusters In the Zone of Avoidance \citep[CIZA,][]{ebeling2002}), and from ROSAT serendipitous catalogues (i.e., the 160 square degree ROSAT Survey catalogue \citep[160SD,][]{mullis2003}, the 400 square degree ROSAT Cluster Survey catalogue \citep[400SD,][]{burenin2007}, the bright SHARC survey cluster catalogue \citep{romer2000}, the Southern SHARC catalogue \citep{burke2003}, the WARPS survey catalogues \citep{perlman2002, horner2008}, and the Einstein Extended Medium Sensitivity Survey catalogue \citep[EMSS,][]{gioia1990}). The MCXC provides a mass $\mathrm{M}_{500}$, a radius $\mathrm{R}_{500}$, and a redshift $z$ for 1,743 galaxy clusters in the all sky. The mass and redshift distributions of MCXC clusters are shown and compared to other SZ and optical catalogues in Fig.~\ref{cluster_prop}. The MCXC contains mainly lower mass clusters than the PSZ2 or the RedMaPPer cluster catalogues.

\subsection{RedMaPPer}

The Red-sequence Matched-filter Probabilistic Percolation \citep[RedMaPPer,][]{rykoff2014} is an algorithm developed to detect clusters in large galaxy optical surveys, such as the SDSS or the Dark Energy Survey\footnote{\url{https://www.darkenergysurvey.org}} \citep[DES,][]{des2005}. Based on the detection of red-sequence-galaxy over-densities, the algorithm provides positions and redshift probability distributions for the detected clusters, together with membership probabilities assigned to galaxies, and a richness $\lambda$ related to the number of galaxies in the clusters. \cite{rykoff2014} have successfully applied RedMaPPer to the SDSS DR8 spectroscopic galaxies and have detected 25,325 galaxy clusters in the redshift range $0.08 < z < 0.55$, over approximately 10,500 squared degrees on the sky. The RedMaPPer catalogue has been extensively studied in different wavelengths \citep[e.g.,][]{saro2015, hurier2018_redm, geach2017}, allowing the confirmations of the galaxy clusters and the measurements of their properties (e.g., their masses). The resdhift $z$ and the mass $\mathrm{M}_{500}$ distributions are shown in Fig.~\ref{cluster_prop}, compared with other catalogues of galaxy clusters detected in different wavelengths that are presented hereafter. For this figure, the scaling relation between the richness and the stellar mass $\mathrm{M}_{500}$ from \cite{saro2015} has been used to compute the mass estimation of the clusters. 

\begin{table*}
    \centering
    \begin{tabular}{c c c c}
        Catalogues & Number of sources & Description \\ \hline\hline
         \texttt{Planck\_z} & 1,094 & \textit{Planck} confirmed clusters\\
         \texttt{Planck\_no-z} & 559 & \textit{Planck} unconfirmed clusters\\
         \texttt{MCXCwP} & 1,193 & MCXC clusters without all \textit{Planck} sources\\
         \texttt{RM}$_{50}$ & 5,067 & RedMaPPer clusters with $\lambda > 50$ without all \textit{Planck} and MCXC sources\\
         \texttt{RM}$_{30}$ & 14,956 & RedMaPPer clusters with $\lambda > 30$ without all \textit{Planck} and MCXC sources\\
    \end{tabular}
    \caption{Summary of the catalogues used in this study. This shows the number of sources and a short description of the five catalogues used to train the U-Net model: the \texttt{Planck\_z}, the \texttt{Planck\_no-z}, the \texttt{MCXCwP}, the \texttt{RM}$_{50}$, and the \texttt{RM}$_{30}$ catalogues.}
    \label{tab:catalogues}
\end{table*}{}

\begin{figure*}[!ht]
\centering
\includegraphics[width=0.5\textwidth]{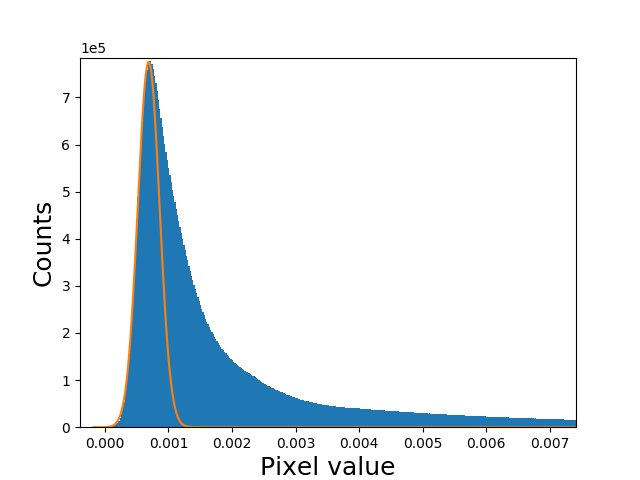}\includegraphics[width=0.5\textwidth]{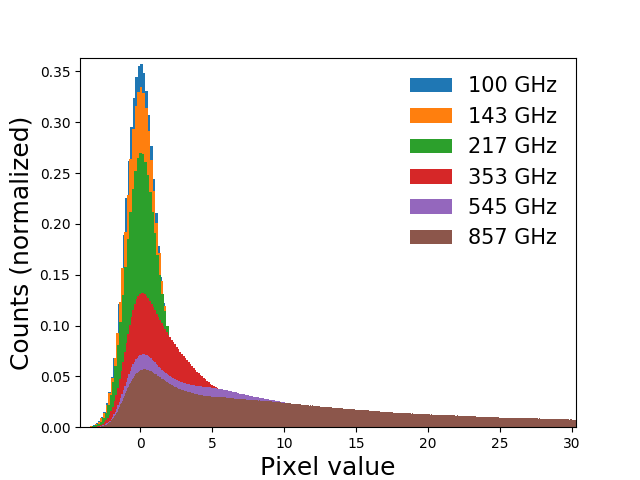}
\caption{\label{norm_data}Illustration of the data pre-processing. Left: the pixel distribution of the map at 353 GHz. A Gaussian is fitted in orange up to the statistical mode of the distribution. The mean and standard deviation of the fitted Gaussian are used to normalise the data. Right: pixel distribution after normalisation of the six \textit{Planck} HFI frequency maps.}
\end{figure*}

\section{Learning procedure}\label{sect:methods}

In this study, we have trained a deep learning algorithm applied on the \textit{Planck} HFI frequency maps to detect low signal-to-noise SZ emission via high signal-to-noise SZ emission coming from the hot gas in galaxy clusters. To do so, we have chosen as inputs of the machine learning algorithm small patches of the \textit{Planck} HFI frequency maps, and as outputs, segmentation maps, showing the positions of the clusters in the patches. The trained model thus provides an SZ prediction map, between 0 and 1, that can be compared with known clusters in a test sample, or with the \textit{Planck} MILCA SZ map.

\subsection{Training catalogue}

\subsubsection{Catalogue of clusters}

We have selected three catalogues of galaxy clusters to construct the segmentation maps that are used as output data for the training of the machine learning algorithm. First, the PSZ2 catalogue of clusters, to start with the very simplest case: learning \textit{Planck} with \textit{Planck}. To ensure the purity of the catalogue, we have selected the 1,094 PSZ2 sources that are confirmed galaxy clusters, i.e., with measured redshifts. We note this cleaned catalogue the \texttt{Planck\_z} catalogue. We note the catalogue of remaining candidates the \texttt{Planck\_no-z} catalogue. Second, we have chosen the MCXC catalogue of X-ray clusters, as it contains lower mass clusters. We have selected MCXC clusters that are not included in the \textit{Planck} catalogue to construct a catalogue with less massive clusters. In the following, we note the 1,193 galaxy clusters of the MCXC catalogue that are not included in the PSZ2 catalogue, the \texttt{MCXCwP} catalogue. Finally, we have selected clusters from the RedMaPPer cluster catalogue from optical data to test the limits of the model and try to detect very low SZ signals. This catalogue contains lower mass and higher redshift clusters. We have used RedMaPPer with different selections in richness $\lambda$ (relative to the number of galaxies in the clusters), i.e., different selection in mass. We note \texttt{RM}$_{i}$ the selection of the RedMaPPer clusters with the criterion: $\lambda > i$. In the following, we use the \texttt{RM}$_{50}$ and the \texttt{RM}$_{30}$ cluster catalogues. We show in Tab.~\ref{tab:catalogues} a summary of the different catalogues presented here above and used in this study.

\subsubsection{Training set and test set}\label{sect:traintest}

We have used HEALPIX with $n_\mathrm{side}=2$ to split the sky onto 48 tiles of equal-sized area of 860 square degrees each. One of them, the seventh one, is arbitrarily chosen to define a test area. This area is centred on the position $(l=112.5^\circ, b=41.81^\circ)$. In this area, there are 40 clusters from the \texttt{Planck\_z} catalogue, 18 from the \texttt{Planck\_no-z} catalogue, and 50 from the \texttt{MCXCwP} catalogue. This area, and especially the clusters inside, are used as a test area and a test sample, and none of its 860 square degrees is seen by the model during the training. The training set is based on patches (projected as described in the next section), extracted in the remaining 47 tiles of the sky.

\subsubsection{\textit{Planck} patches and segmentations}

We have extracted from the \textit{Planck} HFI frequency maps $n=100,000$ multi-channel patches, of $64\times64$ pixels with a resolution of $\theta_{\mathrm{pix}}$=1.7 arcmin (giving a field of view of $1.83^\circ \times 1.83^\circ$). These patches are chosen in the sky with random positions and random orientations, but each of the 100,000 patches contains at least one galaxy cluster of the cluster catalogue chosen as output. {To construct a training sample as well as a validation sample that must be independent, 10,000 patches over the entire sample of 100,000 patches were chosen so that their positions fall into the pixels number 10, 39, and 42 on the HEALPIX map with $n_\mathrm{side}=2$ that split the sky onto 48 tiles of equal-sized area of 860 square degrees (same method as for the construction of the test area). The positions of the validation sample have been split into three regions (i.e., pixels 10, 39, and 42) in the sky, to take into account the noise in the \textit{Planck} maps that is not uniform. This gives 10\% of the entire sample that is set aside for the validation sample, and for which the regions are not seen by the U-Net during the training process.} We have then constructed the segmentation maps associated with the 100,000 patches, by drawing circles at the positions of the clusters. The pixels showing the positions of the clusters are set to 1, while the pixels in any other regions are set to 0. The diameter of the circles showing the positions of the galaxy clusters is arbitrarily set to 5 arcmin, i.e., the size of the smallest beam in the \textit{Planck} HFI frequency maps (at 857 GHz). This fixed size acts like a filter and probably induces a bias in the reconstruction of the SZ sources in the SZ prediction maps, preventing us from computing any reliable flux. However, the information up to a resolution of 5 arcmin can be learned by the network. We note that changing this diameter may change the results but we have not studied this effect in this proof-of-concept paper.

Finally, the dimension of the input data is $100,000\times64\times64\times6$ pixels, and the dimension of the output data is $100,000\times64\times64\times1$ pixels.

\subsection{Data pre-processing}

To successfully apply the deep learning algorithm on \textit{Planck} frequency HFI maps, a pre-processing of the data is needed. The mean input data and their standard deviations should be of the order of the unity, as machine learning algorithms perform better results for this range of values. However, in the \textit{Planck} maps, a large variety of sources are detected, that are producing signals with very different spectral responses (e.g., radio sources bright in the low frequencies, dust sources bright in the high frequencies). The shapes of the pixel distributions of the \textit{Planck} HFI frequency maps are thus very non Gaussian, preventing a simple normalisation of the maps to their means and their standard deviations. Here, we have chosen an approach to optimise the capture of the CMB spectrum deviation at the scale of the CMB fluctuation values (i.e., the secondary anisotropies, in particular the SZ effect).

The primary CMB fluctuations are themselves Gaussian distributed \citep{planck_gaussian2016}, but any kind of external sources (other than CMB anisotropies) add positive emissions in the \textit{Planck} frequency maps. This produces an asymmetric Gaussian distribution, extended to the right part of the pixel distributions. Therefore, we have fitted a Gaussian to the left part of the distributions, up to their statistical modes (values that appear most often), as shown in the left panel of Fig.~\ref{norm_data}. This part of the pixel distributions must contain the noise and the CMB fluctuations in each frequency. We have then normalised each maps by the means and the standard deviations of the fitted Gaussian. This approach optimises the use of deep learning algorithms on the \textit{Planck} frequency maps, especially for the study of the CMB fluctuations at each frequencies and thus of the SZ effect. The pixel distributions of each of the pre-processed HFI frequency maps are shown in the right panel of Fig.~\ref{norm_data}.

\begin{figure*}[!ht]
\centering
\includegraphics[width=0.5\textwidth]{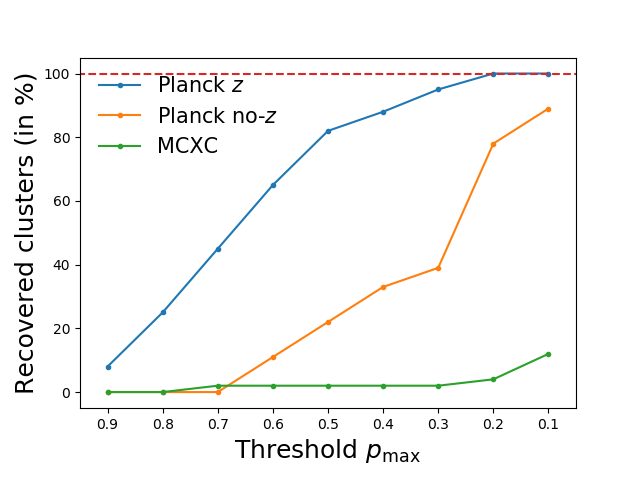}\includegraphics[width=0.5\textwidth]{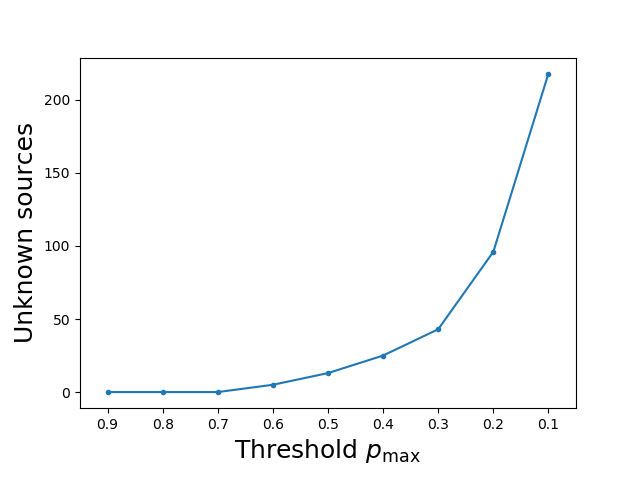}
\caption{\label{complete}Results on the test area containing 40 \texttt{Planck\_z}, 18 \texttt{Planck\_no-z}, and 50 \texttt{MCXCwP} galaxy clusters. Left: galaxy clusters recovered with different threshold of detection $p_\mathrm{max}$. Right: the number of sources recovered with the U-Net not belonging to the \textit{Planck} or the MCXC catalogue as a function of the threshold $p_\mathrm{max}$.}
\end{figure*}

\begin{figure*}[!ht]
\hspace{-0.5cm}\includegraphics[width=1.05\textwidth]{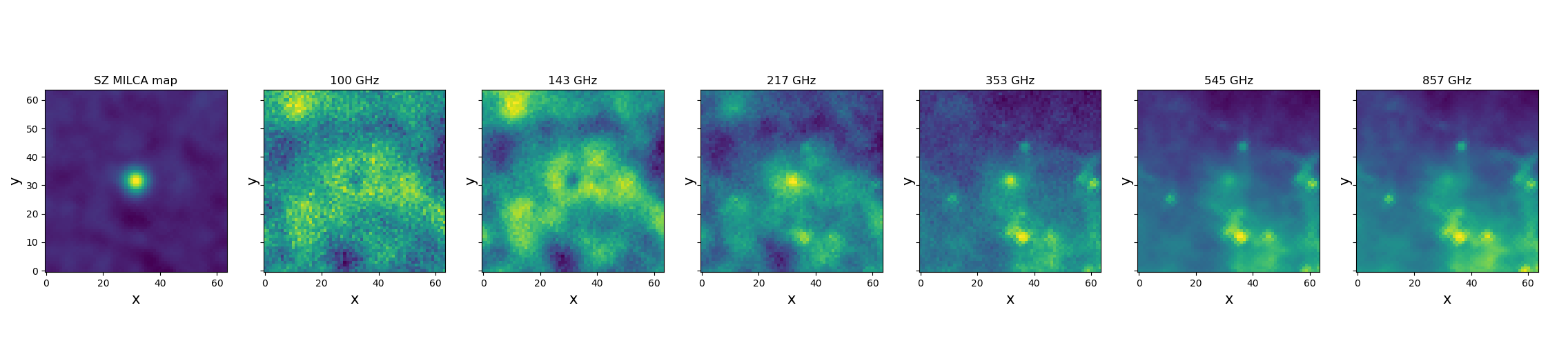}
\caption{\label{new_stack}From left to right, stack of the SZ map and of the six HFI frequency maps at the positions of the 218 sources detected with the U-Net and not belonging to the \textit{Planck} or the MCXC catalogue in the test area. The statistical signature of the SZ effect is suggested in the centres of the HFI frequency maps, together with a complex combination of infra-red sources and CMB structures.}
\end{figure*}

\begin{figure*}[!ht]
\centering
\includegraphics[width=\textwidth]{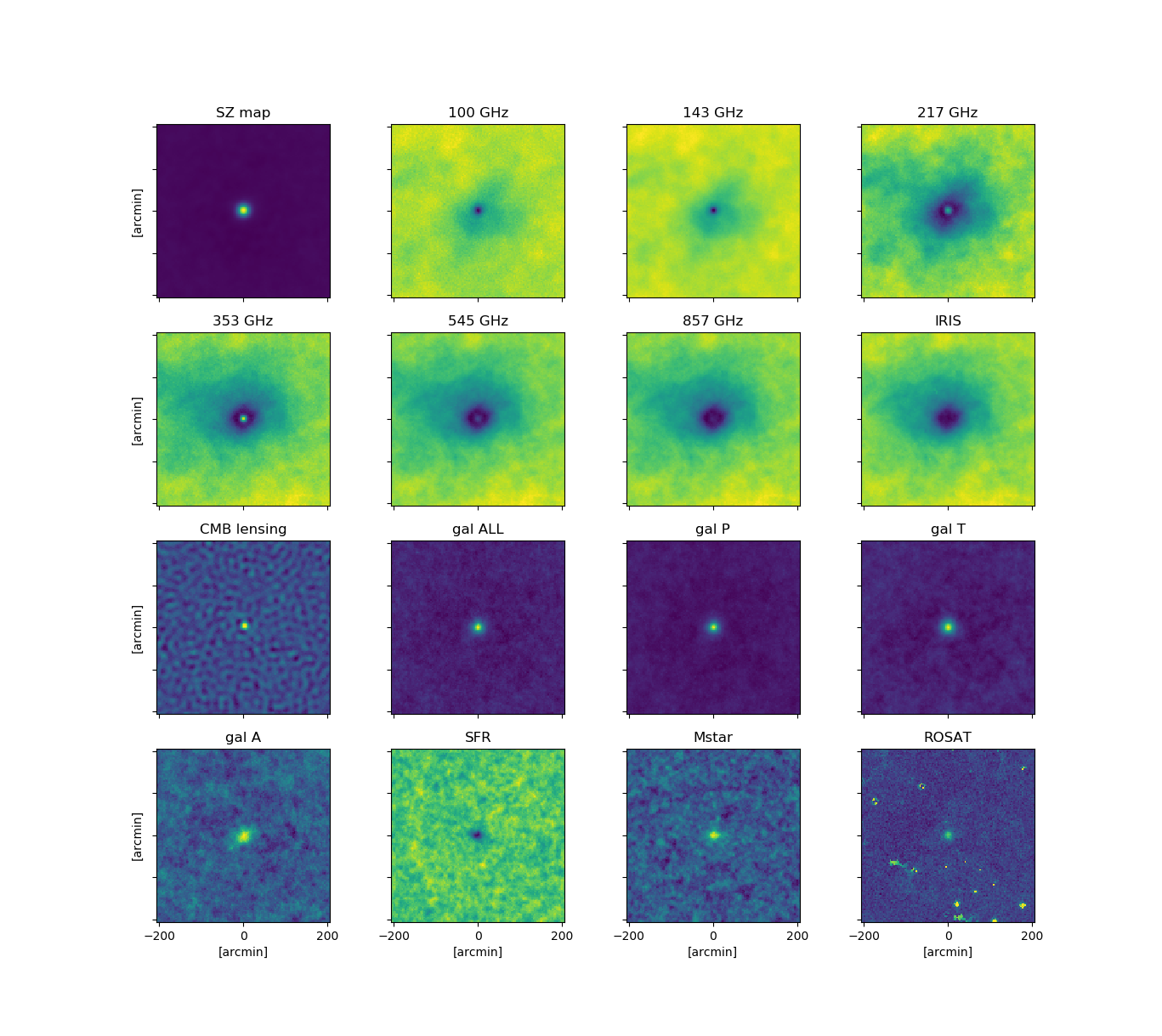}
\caption{\label{full_stack}Stack of the 18,415 sources detected with the U-Net and not belonging to the \textit{Planck} or the MCXC catalogue in 16 different maps, for the chasing of statistical counterparts of galaxy cluster components in different wavelengths: the SZ MILCA $y$ map, the six \textit{Planck} HFI frequency maps, the IRIS map at 100 $\mu$m, the CMB lensing map, the four galaxy over-density maps based on all, and on passive, transitioning, and active populations of galaxies from the WISExSCOS photometric redshift catalogue (in the redshift range $0.1<z<0.3$), and the ROSAT X-ray map. Each of them show statistical hints of galaxy cluster emissions. Patches are $3.4^\circ \times 3.4^\circ$ in order to see the large scale contribution around the sources.}
\end{figure*}

\subsection{U-Net architecture}

In this study, we train a Convolutional Neural Networks (CNN) applied on the Planck HFI frequency maps to detect low signal-to-noise SZ sources via high signal-to-noise SZ emissions coming from the hot gas inside the galaxy clusters. CNN are indeed very efficient to encode information on extended objects that can be invariant in translation or in rotation, such as the case for SZ sources in the \textit{Planck} maps. Moreover, the U-Net architecture, based on CNN, is one of the most efficient to reconstruct segmentation images \citep{ronneberger2015} (e.g., the U-Net architecture has won the ISBI cell tracking challenge 2015 applied on biomedical image segmentation). For these reasons, we have naturally chosen this network architecture to perform the training. We train the U-Net by choosing as inputs small patches of the Planck HFI frequency maps, and as outputs segmentation maps showing the positions of the clusters in the patches. The model will thus provide an SZ prediction map, between 0 and 1, that we can compare with the known clusters in a test sample, or with the Planck MILCA SZ map.

The architecture of the U-Net is symmetric and is composed of two parts: a contracting part and an expansive part, that encode and decode the information. The encoder is composed of different layers of convolutions, that filters the input images with convolutional kernels. The weights of the kernels are learned during the training, so that the encoder can represents itself the information contains on the data in a reduced and optimal dimensional space. The scales of the captured features are increasing with the layers. The second block, or decoder, deconvolve the convolved maps and communicate with the smaller scales to reconstruct forms or objects. Its architecture makes this network very efficient to identify objects and to reconstruct them and classify them in segmentation maps. This network has already been successfully applied in astrophysics for example to mimic numerical simulations, detect cosmic structures, or compute fluxes in blended galaxies \citep[e.g.,][]{he2018, aragon2019, boucaud2019}.

The different blocks of the encoder in the U-Net are based on a typical convolutional network architecture: two $3\times3$ convolution layers, followed by a Rectified Linear Unit (ReLU) activation function $\phi$ that quantifies the significance of the information learned by the kernels:

\begin{equation}
\phi(x) = \left\{
    \begin{array}{ll}
        x & \mbox{if $x \geq 0$,} \\
        0 & \mbox{if $x < 0$.}
    \end{array}
\right.   
\end{equation}

Then, a max pooling layer (or downsampling) is added to reduce the dimensions of the convolved images. The blocks are repeated five times, with the number of filters doubled at each downsampling, as the complexity of the possible features increases with the decreasing scales. The convolved images in the output of the blocks are the input images of the next blocks. The decoder is also composed of a succession of blocks: a $2\times2$ upconvolution, with the number of filters decreasing at each blocks, a concatenation with the convolved images of the same dimension from the encoder part, followed by two $3\times3$ convolutions, with a ReLU activation function. A last $1\times1$ convolution layer is added at the end of the network for a total of 23 layers. We also have added a dropout of 0.2 after each convolutional layers, that helps the network to learn more efficiently the most relevant information in the different filters. The number of filters in the beginning of the network will be related to the complexity of the recovered different features. As the maps of Planck we used does not contain much complex spatial features, we have started the number of kernels at 8, increasing to 128 in the last layer of the encoder. As a last activation function, we use a sigmoid function $s$ to convert the maps at the output of the network into a probability map, bounded between 0 and 1:

\begin{equation}
s(x)=\frac{1}{1+e^{-x}}.
\end{equation}{}

We use the computationally efficient Adam optimizer to update the network weights, with a learning rate of $lr=10^{-4}$ to adapt iteratively the precision of the back-propagated errors. We use a binary cross-entropy loss function to compare the errors between the prediction maps output by the network and the segmentation maps used as outputs. Training has been performed with a batch size of 20, and a patience condition of 12 epochs without improvement of the value of the loss (computed on the validation set not used in the training) before stopping the training. All the parameters on which the training depend (i.e., dropout rate, kernel size, number of filters, initial learning rate) were varied in different trainings of the U-Net, until finding the one actually learning the SZ features.

\begin{figure}[!ht]
\centering
\includegraphics[width=0.5\textwidth]{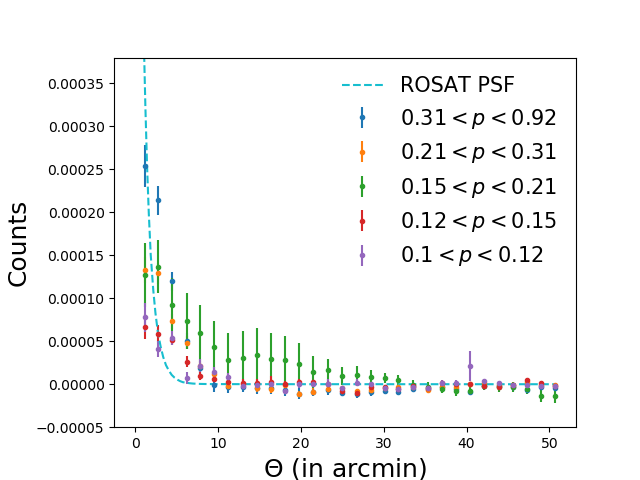}
\caption{\label{x_cool}Stacked radial profiles of the 18,415 sources detected with the U-Net in the X-ray ROSAT map as a function of their associated prediction indexes $p$. The background of the 18,415 maps have been substracted individually to take care of the galactic plane. The cyan dotted line indicate the Point Spread Function (PSF) of ROSAT, normalised to the profile of the first bin. This suggests statistical hints of diffuse gas emission in X-ray for the 18,415 detections, that is decreasing as a function of their prediction indexes output by the U-Net. Error bars computed with the bootstrap method ensures the significance by re-sampling the profiles.}
\end{figure}

\section{Results}\label{sect:results}

\begin{figure}[!ht]
\centering
\includegraphics[width=0.5\textwidth]{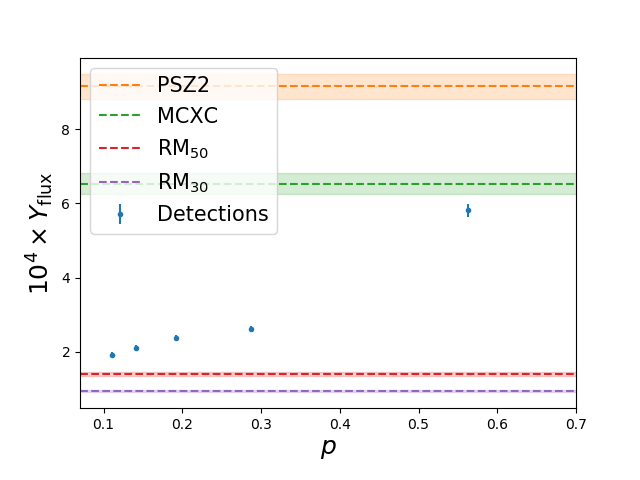}
\caption{\label{scaling}Scaling relation between the prediction index output by the U-Net $p$ and the flux in the MILCA $y$ map. In blue, the fluxes of the 20,204 sources detected on the full sky with the U-Net, stacked on the SZ MILCA $y$ map as a function of their associated prediction index $p$. For comparison, the average flux in the MILCA $y$ map of the PSZ2 clusters is shown in orange, the average flux of the MCXC clusters is shown in green, and the average fluxes of the \texttt{RM}$_{50}$ and the \texttt{RM}$_{30}$ clusters are shown in red and purple, respectively. Error-bars are computed with bootstrap.}
\end{figure}

We have investigated two cases. First, by training with the \texttt{Planck\_z} clusters, we have studied the possibility of potentially detecting more clusters in the \textit{Planck} maps with deep learning. We show the results hereafter. Second, we have trained three other models based on four different cluster catalogues in the output segmentation maps: we have added successively \texttt{MCXCwP}, \texttt{RM}$_{50}$, and \texttt{RM}$_{30}$. Those clusters, even though not included in the \textit{Planck} catalogue, show statistical evidence of SZ emission (as seen by the positive fluxes in Fig.~\ref{scaling} that is described later). This was performed to study the possibility of detecting very diffuse SZ emission, and reconstructing an SZ prediction map. The results are shown in Sect.~\ref{sect:whim}

\subsection{Results on the test area}\label{sect:pl}

To train the algorithm to detect low signal-to-noise SZ sources in \textit{Planck} HFI frequency maps, we have started with the simplest possible case: learn \textit{Planck} with \textit{Planck}. The segmentation maps used as outputs were generated based on the \texttt{Planck\_z} catalogue, to ensure an absolutely pure catalogue of galaxy clusters. The U-net has been successfully trained for about $\sim$3 h on a GPU NVidia Tesla K80, with more than 30 epochs. Based on the trained model, full sky map of SZ prediction $p$ have been constructed.

To estimate the performance of the model and ensure that the U-Net has learned to detect SZ sources, we have compared the SZ prediction map in the test area and the test catalogues described in Sect.~\ref{sect:traintest}. To detect galaxy clusters, we have simply defined the clusters as areas of prediction index $p$ greater than a threshold $p_\mathrm{max}$. For each area recovered above $p_\mathrm{max}$, we have computed the position as the barycentre of the pixels. This detection method is very simplistic and not optimal but it is yet efficient enough to roughly check overall consistency. We have cross-matched the sources detected with this method with the three catalogues: \texttt{Planck\_z}, \texttt{Planck\_no-z}, and \texttt{MCXCwP}, and have studied the recovered clusters as a function of the detection threshold $p_\mathrm{max}$. The recovered clusters on the three catalogues are shown in the left panel of Fig.~\ref{complete}, together with the detected sources, potentially SZ source candidates, that are not in the \textit{Planck} catalogue or in the MCXC catalogue in the right panel. For a threshold $p_\mathrm{max}=0.1$, all the \texttt{Planck\_z} clusters are recovered, together with 89\% of \texttt{Planck\_no-z} clusters, and 12\% of \texttt{MCXCwP} clusters. For the same threshold, 218 sources do not belong to any of the mentioned catalogues.

We have investigated the nature of the 218 sources detected in the test area with the threshold $p_\mathrm{max}=0.1$ and not belonging to any mentioned catalogue, by stacking them in the \textit{Planck} MILCA SZ map and in the \textit{Planck} HFI frequency maps. We are chasing hints of galaxy clusters signatures and want to ensure that they do not statistically correspond to obvious point sources that might contaminate the model (e.g., infra-red point sources). The results of the stack are shown in Fig.~\ref{new_stack}. A statistical presence of SZ emission is suggested by a $y$ emission in the \textit{Planck} MILCA SZ map and by a signature of the SZ effect in the centre of the HFI frequency maps are seen (i.e., a negative emission in the 100 and 143 GHz maps, and a positive emission for the frequencies above 217 GHz). SZ sources may be populated by dust, suggested by the excess of signal also seen in the centre of the map at 217 GHz. A bright infra-red source is also seen in the HFI frequency stacked maps (in the lower part of the quadrants), with an intensity increasing with the frequencies, together with a complex background in the 100, 143, and 217 GHz stacked maps, mainly coming from the CMB.

\begin{figure*}[!ht]
\centering
\includegraphics[width=\textwidth]{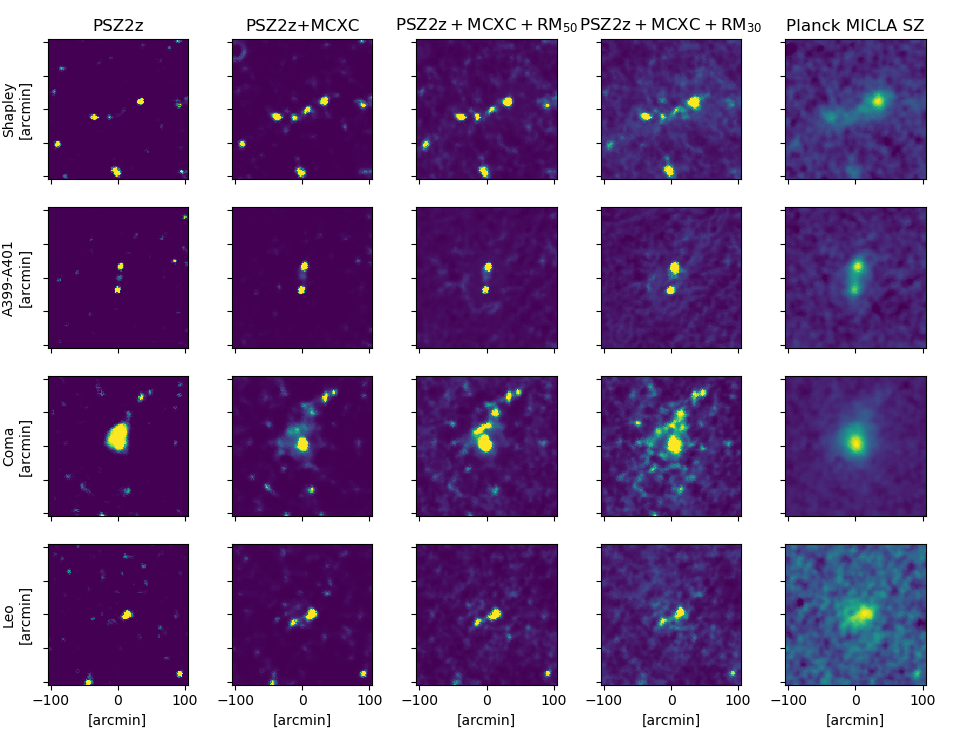}
\caption{\label{patch}Patches of $120\times120$ pixels of size $\theta_{\mathrm{pix}}$=1.7 arcmin in four models of learning and in the SZ MILCA $y$ map, for four emblematic large scale structures. From top to bottom: the Shapley super-cluster, the galaxy cluster pair A399-A401, the Coma super-cluster, and the Leo cluster. From left to right: the model PSZ2z, PSZ2Z+MCXC, PSZ2z+MCXC+RM50, PSZ2Z+MCXC+RM30, and the SZ MILCA $y$ map for comparison.}
\end{figure*}

\subsection{Results on the full sky}

After checking the results of the U-Net in the test area, we have applied the same detection method to the full-sky SZ prediction map, with a detection threshold of $p_\mathrm{max}=0.1$ in order to recover the maximum number of \texttt{Planck\_z} clusters. 20,204 sources are detected in the full sky map with the U-Net with $p_\mathrm{max}=0.1$. We have compared the detections with the three catalogues of known galaxy clusters: \texttt{Planck\_z}, \texttt{Planck\_no-z}, and \texttt{MCXCwP}. Among the 20,204 detected sources, 98.5\% of the \texttt{Planck\_z} clusters are recovered, together with 76.4\% of \texttt{Planck\_no-z} clusters, and 20.8\% of \texttt{MCXCwP} clusters. There are also 11 cluster identified by the Atacama Cosmology Telescope (ACT) \citep{hasselfield2013} and 98 clusters identified by the South Pole Telescope (SPT) \citep{bleem2015} but not included in the \textit{Planck} PSZ2 catalogue. Thus, 18,415 sources do not belong to any of the mentioned catalogues. We have investigated the nature of the sources detected with the U-Net. First, we have cross-matched the sample of 18,415 sources with \textit{Planck} point sources. Only 6.1\% are matched within a cross-match radius of 5 arcmin with the positions of the \textit{Planck} catalogue of galactic cold cores, and only 0.2\% are matched with the positions of the \textit{Planck} sources identified at 353 GHz. Second, we have stacked at their positions 16 maps in different wavelengths, each of them potentially probing different galaxy cluster counterparts. Some of the 16 maps are also based on \textit{Planck} data, and thus are not independent, but some of the maps are independent and may show hints of galaxy cluster counterparts in other wavelengths, i.e., in near infra-red (where galaxies emit) and in X-rays (where the same gas as the one seen in the SZ emits). The 16 maps are the \textit{Planck} SZ MILCA map, the six \textit{Planck} HFI frequency maps, the IRIS map at 100 $\mu$m \citep{mamd2005}, the CMB lensing map (based on \textit{Planck}) \citep{planck_cmb_lensing}, four galaxy over-density maps of all galaxies (noted GAL ALL), passive galaxies (noted GAL P), transitioning galaxies (noted GAL T), and active galaxies (noted GAL A) following \cite{bonjean2019}, a star formation rate density maps (noted SFR) and a stellar mass density maps (noted Mstar) with the method from \cite{bonjean2019}, and finally, the ROSAT X-ray map (ByoPiC\footnote{\url{https://byopic.eu}} product). The \textit{Planck} maps are masked from the \textit{Planck} Catalogue of Compact Sources \citep[PCCS,][]{planck_pccs}, and the ROSAT map is masked from the point sources detected in ROSAT, Chandra, and XMM-Newton \citep[][ respectively]{boller2016, evans2010, rosen2016}. The result of the 16 stacks are shown in Fig.~\ref{full_stack}.

The stacks show hints of statistical emission of galaxy clusters. A statistical hint of dark matter potential is seen in the CMB lensing map. In the over-density maps and galaxy properties density maps, a statistical hint of presence of passive galaxy over-density is seen in each of the maps, suggesting that some of the detected SZ sources might be galaxy clusters in the redshift range $0.1<z<0.3$ (that is the redshift range of the catalogue of galaxies used to construct the galaxy density maps). In the six \textit{Planck} HFI frequency maps, a complex background is seen, that correlates with the IRIS map at 100$\mu$m showing the dust. This suggests that the background in the \textit{Planck} frequency maps is due to the emission of the dust in our galaxy, and thus that the sources detected with the U-Net are preferentially in areas without this foreground dust contamination. In the very centre of the HFI maps, inside the hole produced by the background, a statistical SZ signature is seen, with a decrement at 100 and 143 GHZ, and an increment in the 353 GHZ map, decreasing up to the 857 GHz map. A statistical hint of Bremsstrahlung emission is also seen in the ROSAT stacked map. Each of these stacks shows that there are statistical hints of cluster counterparts in different wavelengths, and thus that potentially some of the sources detected with the U-Net might be actual galaxy clusters. To ensure that setting the threshold $p_\mathrm{max}$ to 0.1 is acceptable, we have also stacked the sources detected with the U-Net in bins of their maximum associated prediction index $p$, in the X-ray ROSAT map. The bins are chosen so that they contain the same number of sources, i.e., 3,683 each. The stacked radial profiles measured in the ROSAT map are shown in Fig.~\ref{x_cool}, with error bars computed with the bootstrap method that re-samples the profiles in each bin. The background of each of the 18,415 sources detected with the U-Net and not belonging to the \textit{Planck} or the MCMC catalogues have been substracted on their maps individually to take count of the galactic plane. Figure \ref{x_cool} also shows the Point Spread Function (PSF) of ROSAT with an effective beam of about 2 arcminutes, normalised to the profile of the first bin. 

This result shows statistical hint of X-ray emission potentially coming from actual galaxy clusters for each bin of associated prediction index $p$, and decreasing with $p$. Moreover, X-ray emission detected for the five bins are diffuse emission, as seen by the radial extension of the profiles that are larger than the ROSAT PSF. The bootstrap method ensures the significance of the profiles by re-sampling the profiles in each bin. Even in the lowest prediction index bin, $0.1<p<0.12$, a diffuse X-ray emission is observed. This suggests that the low prediction index output by the U-Net may indicate potential low signal-to-noise ratio SZ emissions, e.g., low pressure, low mass, or high $z$ galaxy clusters.

\subsection{Scaling relation}

These results show that the U-Net trained on high signal-to-noise ratio SZ sources (i.e., \textit{Planck} clusters) has learned the frequency-dependency and the spatial features of the SZ effect in the six \textit{Planck} HFI frequency maps, and that potentially some of the sources detected with the U-Net and not belonging to the \textit{Planck} or the MCXC catalogue might be actual galaxy clusters, even at the lowest bin of associated prediction index $p$. Results shown in the stacked profiles of the X-Ray maps show that it seems to have a correlation between the amplitude of the signal and $p$. If so, $p$ should be somehow related to the SZ flux in the MILCA $y$ map. We have analysed the associated prediction index $p$ of the 20,204 sources detected with the U-Net in the full-sky, by stacking at their positions the \textit{Planck} SZ MILCA map for different bins of  $p$. Results are shown in Fig.~\ref{scaling}, where the blue points are the fluxes (computed with aperture photometry) in the stacked \textit{Planck} SZ MILCA maps as a function of the mean prediction index $p$ in the bins. The stacked flux (also computed by aperture photometry) in the \textit{Planck} SZ MILCA map of the \textit{Planck} PSZ2 clusters is displayed in orange for comparison, together with the stacked flux of the MCXC clusters in green, and the stacked fluxes of the RedMaPPer clusters selected in richness in red and purple. All the errors are computed with the bootstrap method. This result suggests that the maximum value of prediction index $p$ output by the U-Net might be statistically related to the integrated flux in the \textit{Planck} SZ MILCA map. Therefore, in the future and with more investigation, a scaling law may be estimated to translate the maps from $p$ units to $y$ units.

In Fig.~\ref{scaling}, we also show that the sources detected with the U-Net have average fluxes between the ones of the MCXC and the RedMaPPer clusters. This suggest that the sources detected with the U-Net, if actual clusters, may contain less massive SZ sources at low redshift, or higher redshift SZ sources, so that the stellar mass and redshift distributions of the sources detected with the U-Net are between the distributions of the MCXC and the RedMaPPer clusters that are shown in Fig.~\ref{cluster_prop}. However, as the SZ effect is a projected effect, there is a degeneracy between the redshift and the mass of the individual clusters. A detailed investigation of the range of redshifts and masses of the new detected sources is not performed in the present study, that only shows the potential of the application of deep learning algorithms on the \textit{Planck} data.

\subsection{Diffuse SZ emission}\label{sect:whim}

Motivated by the construction of a new SZ map more sensitive to lower signal-to-noise SZ emission, we have trained three other U-nets, by choosing lower SZ signal-to-noise ratio galaxy cluster catalogues of reference to construct the segmentation maps of the training catalogue. In practice, we have successively added the \texttt{MCXCwP} clusters, the \texttt{RM}$_{50}$, and finally the \texttt{RM}$_{30}$ clusters, in addition to the \texttt{Planck\_z} clusters. There are in total four U-Net models (the first one being the one presented in the previous section).

We have generated four SZ prediction full-sky maps based these four U-Net trained models. To illustrate the potential recognition of diffuse gas, we have arbitrarily focused on four regions around large-scale structures already identified as containing diffuse SZ signal: the Shapley super-cluster, the galaxy cluster pair A399-A401, the Coma super-cluster, and the Leo super-cluster. Patches extracted from the SZ prediction maps derived from the four models around these structures are shown in Fig.~\ref{patch}, together with patches of the \textit{Planck} SZ MILCA map for visual comparison.

When adding the \texttt{MCXCwP} clusters, the diffuse gas around the super-clusters and the bridge of matter between A399 and A401 are recovered. For the models with the RedMaPPer clusters, potential indications of large scale structures connecting the structures are seen. The SZ prediction maps obtained with the U-Net are visually very close to the MILCA SZ map, but they seem less noisy and better resolved. These preliminary results are promising for the detection and the characterisation of the diffuse gas in the large-scale structures.

\section{Discussion and Summary}\label{sect:discussion}

A U-Net trained on the \textit{Planck} HFI frequency maps at recognising the spatial and spectral features of high signal-to-noise SZ signatures produced by the hot gas in known galaxy clusters is able to recognise \textit{Planck} clusters and shows promising results for the potential detection of lower signal-to-noise SZ sources. In the most conservative case when training with the \texttt{Planck\_z} catalogue of 1,094 sources, 200 MCXC clusters that are not in the \textit{Planck} catalogue are recovered, together with more than 18,000 new potential sources detected with the U-Net above a detection threshold $p_\mathrm{max}=0.1$. This confirm results obtained in \cite{hurier2017} and \cite{tarrio2019}, where they detected about two times more galaxy clusters than in the \textit{Planck} catalogue using \textit{Planck} data. Although the detection threshold has been set to a low value, the different stacks of the 18,415 sources detected with the U-Net in bins of their associated SZ prediction index shows hints of galaxy cluster signatures. The statistical presence of dark matter halos is suggested by the stacked CMB lensing map, the statistical presence of hot gas is suggested by the stacked \textit{Planck} SZ MILCA map and by the X-ray ROSAT map, and the statistical presence of passive galaxy over-density is suggested by stacked galaxy density maps (constructed with the value-added WISExSCOS photometric redshift catalogue). With more investigation, in the future, a catalogue of potential galaxy cluster candidates might be constructed thanks to this approach. Such potential candidates might be compared to the next generation of full sky surveys allowing the confirmation of the presence of diffuse gas (i.e., with SRG/eROSITA in the X-rays). Galaxy clusters or complex structures that might be newly detected with this method are expected at intermediate or high redshift (i.e., $z\geq0.5$).

By focusing on areas around multiple-cluster systems, we have shown that deep learning models can be used to reconstruct an SZ map, more sensitive to lower signal-to-noise ratio SZ emissions. Although a qualitative study is still to perform in a future analysis, this proof-of-concept study shows the potential of applying deep learning algorithms on \textit{Planck} data.

Furthermore, the U-Net and the method presented in this study can in principle also be applied to any components separation in the \textit{Planck} data, e.g., radio emission, dust emission, CO emission, CMB, and so on. We are presently investigating the training of more complex deep learning architectures multi-classes classification in order to perform multi-component separation.

To improve even more the results, ACT and SPT higher-resolution maps could be included in the training process: results could then be compared to other studies that combined these two missions with the \textit{Planck} data \citep[e.g.,][]{chown2018, aghanim2019}.

\begin{acknowledgements}
The author thanks the anonymous referee for her/his comments that helped to significantly improve the quality of this paper. The author thanks Marc Huertas-Company for fruitful discussions, and for his PhD course in deep learning. The author also thanks Nabila Aghanim, Philippe Salomé, Tony Bonnaire, Alexandre Beelen, Aurélien Decelle, Marian Douspis, and all the members of the ByoPiC team (\url{https://byopic.eu/team}) for useful comments that have helped to increase the quality of the paper. This research has been supported by the funding for the ByoPiC project from the European Research Council (ERC) under the European Union's Horizon 2020 research and innovation programme grant agreement ERC-2015-AdG 695561. This publication made use of the SZ-Cluster Database (\url{http://szcluster-db.ias.u-psud.fr/sitools/client-user/SZCLUSTER\_DATABASE/project-index.html}) operated by the Integrated Data and Operation Centre (IDOC) at the Institut d'Astrophysique Spatiale (IAS) under contract with CNES and CNRS. This study is based on observations obtained with Planck (\url{http://www.esa.int/Planck}), an ESA science mission with instruments and contributions directly funded by ESA Member States, NASA, and Canada.
\end{acknowledgements}

\bibliographystyle{aa}
\bibliography{aanda}

\end{document}